\documentclass[twocolumn,showpacs,preprintnumbers,amsmath,amssymb]{revtex4}
\usepackage{epsfig}

\usepackage{graphicx}
\usepackage{dcolumn}
\usepackage{bm}
\usepackage{amsthm}

\newcommand{\ket}[1]{\mbox{$ | #1 \rangle $}}
\newcommand{\bra}[1]{\mbox{$ \langle #1 | $}}

\begin{document}

\title{On single-photon quantum key distribution in the presence of loss}

\author{Marcos Curty$^{1,2}$, and Tobias Moroder$^{2,3}$}
\affiliation{ 1 Center for Quantum Information and Quantum
Control, Department of Physics and Department of Electrical \&
Computer Engineering, University of Toronto, Toronto, Ontario,
M5S 3G4, Canada \\
2 Institute for Quantum Computing, University of Waterloo, 200
University Avenue West, Waterloo, Ontario, N2L 3G1, Canada \\
3 Quantum Information Theory Group, Institut f\"ur Theoretische
Physik I, and Max-Planck Research Group, Institute of Optics,
Information and Photonics, Universit\"at Erlangen-N\"urnberg,
91058 Erlangen, Germany }

\date{\today}

\begin{abstract}
We investigate two-way and one-way single-photon quantum key
distribution (QKD) protocols in the presence of loss introduced
by the quantum channel. Our analysis is based on a simple
precondition for secure QKD in each case. In particular, the
legitimate users need to prove that there exists no separable
state (in the case of two-way QKD), or that there exists no
quantum state having a symmetric extension (one-way QKD), that is
compatible with the available measurements results. We show that
both criteria can be formulated as a convex optimisation problem
known as a semidefinite program, which can be efficiently solved.
Moreover, we prove that the solution to the dual optimisation
corresponds to the evaluation of an optimal witness operator that
belongs to the minimal verification set of them for the given
two-way (or one-way) QKD protocol. A positive expectation value of
this optimal witness operator states that no secret key can be
distilled from the available measurements results. We apply such
analysis to several well-known single-photon QKD protocols under
losses.
\end{abstract}


\maketitle

\section{INTRODUCTION}

Quantum key distribution (QKD) protocols typically involve a
two-step procedure in order to generate a secret key
\cite{gisin_rev_mod,norbert06}. First, the legitimate users
(Alice and Bob) perform a set of measurements on effective
bipartite quantum states that are distributed to them. As a
result, they end up with a classical joint probability
distribution, that we shall denote as $p(a_i,b_j)\equiv p_{ij}$,
describing their outcomes. The second step consists of a classical
post-processing of the data $p_{ij}$. It requires an
authenticated classical channel, and it includes post-selection
of data, error-correction to reconcile the data, and privacy
amplification to decouple the data from a possible eavesdropper
(Eve) \cite{gisin_rev_mod,norbert06}.

In order to create the correlated data $p_{ij}$, QKD schemes
usually require Alice to prepare some non-orthogonal quantum
states $\ket{\psi_i}$ with a priori probabilities $p_i$ that are
sent to Bob. On the receiving side, Bob measures each received
signal with a {\it positive operator value measure} (POVM)
$\{B_j\}$. Generalising the ideas introduced by Bennett {\it et
al.} in Ref.~\cite{mermin}, the signal preparation process in
this kind of schemes can alternatively be thought of as follows:
Alice produces first bipartite states
$\ket{\psi_{source}}_{AB}=\sum_i \sqrt{p_i}
\ket{\alpha_i}_A\ket{\psi_i}_B$ and, afterwards, she measures the
first subsystem in the orthogonal basis $\ket{\alpha_i}_A$
corresponding to the measurement operators
$A_i=\ket{\alpha_i}_A\bra{\alpha_i}$. This action generates the
signal states $\ket{\psi_i}$ with a priori probabilities $p_i$.
The reduced density matrix of Alice,
$\rho_A=\textrm{Tr}_B(\ket{\psi_{source}}_{AB}\bra{\psi_{source}})$,
depends only on the probabilities $p_i$ and on the overlap of the
signals states $\ket{\psi_i}$. This means, in particular, that
$\rho_A$ is always fixed by the preparation process and cannot be
modified by Eve. In order to include this information in the
measurement process one can add to the observables $\{A_i\otimes
B_j\}$, measured by Alice and Bob, other observables $\{C_k
\otimes\openone\}$ such that the observables $\{C_k\}$ form a
tomographic complete set of Alice's Hilbert space
\cite{curty04suba}. From now on, we will consider that the data
$p_{ij}$ and the POVM $\{A_i\otimes B_j\}$ include also the
observables $\{C_k \otimes\openone\}$.

The classical post-processing of $p_{ij}$ can involve either
two-way or one-way classical communication. Two-way classical
communication protocols can tolerate a higher error rate than
one-way communication techniques \cite{lo03}. On the other hand,
one-way post-processing methods typically allow to derive simpler
unconditional security proofs for QKD than those based on two-way
communication \cite{mayers,shor1,loQIC,kiyo}. In this last
paradigm, two different cases can be considered: {\it Reverse
reconciliation} (RR) refers to communication from Bob to Alice,
and {\it Direct reconciliation} (DR) permits only communication
from Alice to Bob. (See, for instance, Refs.~\cite{gross,heid}.)

An essential question in QKD is to determine whether the
correlated data $p_{ij}$ allow Alice and Bob to generate a secret
key at all during the second phase of QKD. Here we consider the
so-called {\it trusted device scenario}, where Eve cannot modify
the actual detection devices employed by Alice and Bob, as used in
Refs.~\cite{curty04a,curtynorbert}. We assume that the legitimate
users have complete knowledge about their detection devices,
which are fixed by the actual experiment. The case of two-way
classical post-processing has been analysed in
Ref.~\cite{curty04a}, where it was proven that a necessary
precondition for secure two-way QKD is the provable presence of
quantum correlations in $p_{ij}$. That is, it must be possible to
interprete $p_{ij}$, together with the knowledge of the
corresponding observables $\{A_i\otimes B_j\}$, as coming {\it
exclusively} from an entangled state. Otherwise, no secret key
can be distilled from $p_{ij}$. In order to deliver this
entanglement proof any separability criteria (see, for instance,
Ref.~\cite{Separability} and references therein) might be
employed. The important question here is whether the chosen
criterion can provide a necessary and sufficient condition to
detect entanglement even when the knowledge about the quantum
state is not tomographic complete. It was proven in
Ref.~\cite{curty04a} that entanglement witnesses (EWs) fulfill
this condition. An EW is an Hermitian operator $W$ with a positive
expectation value on all separable states
\cite{ew1,terhal,optew,curty04a}. So, if a state $\rho_{AB}$
obeys $\text{Tr}(\rho_{AB}W)<0$, the state $\rho_{AB}$ must be
entangled. With this separability criterion,
Refs.~\cite{curty04a,curty04suba} analysed three well-known
qubit-based QKD schemes, and provided a compact description of a
minimal verification set of EWs ({\it i.e}, one that does not
contain any redundant EW) for the four-state \cite{bennett84a}
and the six-state \cite{bruss98a} QKD protocols, and a reduced
verification set of EWs ({\it i.e.}, one which may still include
some redundant EWs) for the two-state \cite{ben92} QKD scheme,
respectively. These verification sets of EWs allow a systematic
search for quantum correlations in $p_{ij}$. One negative
expectation value of one EW in the set suffices to detect
entanglement. To guarantee that no verifiable entanglement is
present in $p_{ij}$, however, it is necessary to test {\it all}
the members of the set. Unfortunately, to find a minimal
verification set of EWs, even for ideal qubit-based QKD schemes,
is not always an easy task, and it seems to require a whole
independent analysis for each protocol, let alone for higher
dimensional QKD schemes \cite{curty04a,curty04suba}. (See also
Ref.~\cite{jenshyllus}.) Also, one would like to include in the
analysis the attenuation introduced by the quantum channel, not
considered in Refs.~\cite{curty04a,curty04suba}, and which
represents one of the main limitations for optical realisations of
QKD.

One central observation of this paper is very simple, yet
potentially very useful: Given {\it any} qubit-based two-way QKD
scheme, one can search for quantum correlations in $p_{ij}$ by
just applying the {\it positive partial transposition} (PPT)
criterion \cite{ew1,peres} adapted to the case of a quantum state
that cannot be completely reconstructed. This criterion provides
a necessary and sufficient entanglement verification condition
for any qubit-based QKD protocol even in the presence of loss
introduced by the channel, since, in this scenario, only {\it
nonpositive partial transposed} (NPT) entangled states exist.
Moreover, it is rather simple to evaluate in general since it can
be cast into the form of a convex optimisation problem known as
semidefinite program (SDP)
\cite{vandenberghe:1996,vandenberghebook}. Such instances of
convex optimisation problems can be solved efficiently, for
example by means of interior-point methods
\cite{vandenberghe:1996,vandenberghebook}. This means, in
particular, that this criterion can be applied to any qubit-based
QKD scheme in a completely systematic way.

One-way QKD schemes can be analysed as well with SDP techniques.
It was shown in Ref.~\cite{tobione} that a necessary precondition
for one-way QKD with RR (DR) is that Alice and Bob can prove that
there exists no quantum state having a symmetric extension to two
copies of system $A$ (system $B$) that is compatible with the
observed data $p_{ij}$. This kind of states (with symmetric
extensions) have been analysed in detail in
Refs.~\cite{do1,doherty04,doherty05}, where it was proven that
the search for symmetric extensions for a given quantum state can
be stated as a SDP. (See also Refs.~\cite{ter1,sym1}.) Here we
complete the results contained in Ref.~\cite{tobione}, now
presenting specifically the analysis for the case of a lossy
channel.

Both QKD verification criteria mentioned above, based on SDP
techniques, also provide a means to search for witness operators
for a given two-way or one-way QKD protocol in a similar spirit
as in Refs.~\cite{curty04a,curty04suba}. Any SDP has an
associated dual problem that represents also a SDP
\cite{vandenberghe:1996,vandenberghebook}. This dual problem can
be used to obtain a certificate of infeasibility whenever the
primal problem is actually infeasible. Most importantly, it can
be proven that the solution to this dual problem corresponds to
the evaluation of an optimal witness operator, that belongs to the
minimal verification set of them for the given protocol, on the
observed data $p_{ij}$. A positive expectation value of this
optimal witness operator indicates that no secret key can be
distilled from the observed data $p_{ij}$.

The paper is organised as follows. In Sec.~II we introduce the QKD
verification criteria for two-way and one-way QKD in more detail,
and we show how to cast them as primal SDPs. Then, in Sec.~III,
we present the dual problems associated to these primal SDPs, and
we show that the solution to these dual problems corresponds to
evaluating an optimal witness operator on the observed data
$p_{ij}$ for the given protocol. These results are then
illustrated in Sec.~IV, where we investigate in detail the
two-state QKD protocol \cite{ben92} in the presence of loss. The
analysis for other qubit-based QKD schemes is completely
analogous, and we include very briefly the results of our
investigations on other QKD protocols in an Appendix. Finally,
Sec.~V concludes the paper with a summary.

\section{QKD verification criteria}\label{verif_criteria}

Our starting point is the observed joint probability distribution
$p_{ij}$ obtained by Alice and Bob after their measurements
$\{A_i\otimes B_j\}$. This probability distribution defines an
equivalence class $\mathcal{S}$ of quantum states that are
compatible with it,
\begin{equation}
  \label{eq_class}
  \mathcal{S}=\left\{ \rho_{AB}\ |\ \text{Tr}(A_i \otimes B_j\
  \rho_{AB})=p_{ij}, \ \forall i,j \right\}.
\end{equation}
By definition, every $\rho_{AB} \in \mathcal{S}$ can represent
the state shared by Alice and Bob before their measurements.

In single-photon QKD schemes in the presence of loss, any state
$\rho_{AB} \in \mathcal{S}$ can be described on an Hilbert space
$\mathcal{H}^A_2\otimes\mathcal{H}^B_3$, with $\mathcal{H}^A_2$
and $\mathcal{H}^B_3$ denoting, respectively, Alice's and Bob's
Hilbert spaces, and where the subscript indicates the dimension of
the corresponding Hilbert space. To see this, we follow the
signal preparation model introduced previously, where Alice
prepares states $\ket{\psi_{source}}_{AB}=\sum_{i=0}^{N-1}
\sqrt{p_i}
\ket{\alpha_i}_A\ket{\psi_i}_B\in\mathcal{H}^A_N\otimes\mathcal{H}^B_2$,
and, afterwards, she measures the first subsystem in the
orthogonal basis $\ket{\alpha_i}_A$. Using Neumark's theorem
\cite{neumark1,neumark2}, we can alternatively describe the
preparation process as Alice producing first bipartite states on
$\mathcal{H}^A_2\otimes\mathcal{H}^B_2$ and, afterwards, she
measures the first subsystem with a POVM $\{A_i\}_{i=0}^{N-1}$.
(See also Ref.~\cite{note_neumark}.) To include the loss of a
photon in the quantum channel, we simply enlarge Bob's Hilbert
space from $\mathcal{H}^B_2$ to $\mathcal{H}^B_3$ by adding the
vacuum state $\ket{vac}_B$.

\subsection{Two-way QKD}

Let us now consider two-way QKD protocols. Whenever the observed
joint probability distribution $p_{ij}$, together with the
knowledge of the corresponding measurements performed by Alice
and Bob, can be interpreted as coming from a separable state
$\sigma_{sep}$ then no secret key can be distilled from the
observed data \cite{curty04a}. In
$\mathcal{H}^A_2\otimes\mathcal{H}^B_3$ only NPT entangled states
exist and, therefore, a simple necessary and sufficient criterion
to detect entanglement in this scenario is given by the PPT
criterion \cite{ew1,peres}: A state $\rho_{AB} \in
\mathcal{H}^A_2\otimes\mathcal{H}^B_3$ is separable if and only
if its partial transpose $\rho_{AB}^{\Gamma}$ is a positive
operator. Partial transpose means a transpose with respect to one
of the subsystems \cite{partialtrans}. Such a result is generally
not true in higher dimensions.

{\it Observation 1}: Consider a qubit-based QKD scheme in the
presence of loss where Alice and Bob perform local measurements
with POVM elements $A_i$ and $B_j$, respectively, to obtain the
joint probability distribution of the outcomes $p_{ij}$.
Then, the correlations $p_{ij}$ can originate from a separable
state if and only if there exists $\rho_{AB} \in \mathcal{S}$
such as $\rho_{AB}^{\Gamma}\geq{}0$.

{\it Proof}. If $p_{ij}$ can originate from a separable state,
then there exists $\sigma_{sep}$ such as
$\sigma_{sep}\in\mathcal{S}$. Moreover, we have that any
separable state satisfies $\sigma_{sep}^{\Gamma}\geq{}0$. To
prove the other direction, note that if there exists $\rho_{AB}
\in \mathcal{S}$ such that $\rho_{AB}^{\Gamma}\geq{}0$ then, since
$\rho_{AB}\in\mathcal{H}^A_2\otimes\mathcal{H}^B_3$, we find that
$\rho_{AB}$ must be separable \cite{ew1,peres}. $\blacksquare$

To determine whether there exists $\rho_{AB} \in \mathcal{S}$ such
as $\rho_{AB}^{\Gamma}\geq{}0$ can be solved by means of a primal
semidefinite program (SDP). This is a convex optimisation problem
of the following form:
\begin{eqnarray}\label{primalSDP_marcos}
\text{minimise} && c^T {\bf{x}} \\
\nonumber \text{subject to} && F({\bf{x}})=F_0 + \sum_i x_i F_i
\geq 0,
\end{eqnarray}
where the vector ${\bf x}=(x_1, ..., x_t)^T$ represents the
objective variable, the vector $c$ is fixed by the particular
optimisation problem, and where the matrices $F_0$ and $F_i$ are
Hermitian matrices. The goal is to minimise the linear function
$c^T{\bf{x}}$ subjected to the linear matrix inequality (LMI)
constraint $F({\bf{x}}) \geq 0$
\cite{vandenberghe:1996,vandenberghebook}. If the vector $c=0$,
then the optimisation problem given by
Eq.~(\ref{primalSDP_marcos}) reduces to find whether the LMI
constraint can be satisfied for some value of the vector ${\bf x}$
or not. In this case, the SDP is called a {\it feasibility
problem}. Remarkably, SDPs can be solved with arbitrary accuracy
in polynomial time, for example by means of interior-point methods
\cite{vandenberghe:1996,vandenberghebook}.

According to Observation $1$, we can find whether there exists a
separable state that belongs to the equivalence class
$\mathcal{S}$ just by solving the following feasibility problem
\cite{aclar_LMI}:
\begin{eqnarray}\label{primalSDP_two_way}
\text{minimise} && 0 \\
\nonumber \text{subject to} && \rho_{AB}({\bf x}) \in \mathcal{S}, \\
\nonumber && \rho_{AB}({\bf x})\geq{}0, \\
\nonumber && \rho_{AB}^{\Gamma}({\bf x})\geq{}0,
\end{eqnarray}
where the objective variable ${\bf x}$ is used to parametrise the
density operators $\rho_{AB}$. The method used to parametrise
$\rho_{AB}$ is discussed in detail in Sec.~\ref{parametr}.

\subsection{One-way QKD}

One-way RR (DR) QKD schemes require from Alice and Bob to show
that there exists no quantum state $\rho_{AB}\in \mathcal{S}$ with
a symmetric extension to two copies of system $A$ (system $B$)
\cite{tobione}. A state $\rho_{AB}$ is said to have a symmetric
extension to two copies of system $A$ if and only if there exists
a tripartite state $\rho_{ABA^\prime} \geq 0$, with
$\text{Tr}(\rho_{ABA^\prime})=1$, and where $\mathcal{H}^A \simeq
\mathcal{H}^{A^\prime}$, such that \cite{do1}:
\begin{eqnarray}
\textrm{Tr}_{A^\prime}(\rho_{ABA^\prime}) & = & \rho_{AB},\label{eq1}\\
P \rho_{ABA^\prime} P & = & \rho_{ABA^\prime},\label{eq2}
\end{eqnarray}
where the swap operator $P$ satisfies $P \ket{ijk}_{ABA^\prime} =
\ket{kji}_{ABA^\prime}$. This definition can be easily extended
to cover also the case of symmetric extensions of $\rho_{AB}$ to
two copies of system $B$, and also of extensions of $\rho_{AB}$
to more than two copies of system $A$ or of system $B$ \cite{do1}.

To find whether $\rho_{AB}\in\mathcal{S}$ has a symmetric
extension to two copies of system $A$ can be solved with the
following feasibility problem:
\begin{eqnarray}\label{primalSDP_one_way}
\text{minimise} && 0 \\
\nonumber \text{subject to} &&
\rho_{AB}({\bf x}) \in \mathcal{S}, \\
\nonumber && P\rho_{ABA'}({\bf x})P=\rho_{ABA'}({\bf x}), \\
\nonumber && \rm{Tr}_{A'}[\rho_{ABA'}({\bf x})]=\rho_{AB}({\bf x}), \\
\nonumber && \rho_{ABA'}({\bf x})\geq{}0.
\end{eqnarray}
Note that this SDP does not include the constraint $\rho_{AB}({\bf
x}) \geq 0$ because non-negativity of the extension
$\rho_{ABA'}({\bf x})$, together with the condition
$\text{Tr}_{A'}[\rho_{ABA'}({\bf x})]=\rho_{AB}({\bf
  x})$, already implies non-negativity of $\rho_{AB}({\bf x})$. The SDP for
one-way QKD with DR can be obtained in a similar way.

\subsection{Parametrisation of the SDPs}\label{parametr}

To actually implement the SDPs given by
Eq.~(\ref{primalSDP_two_way}) and Eq.~(\ref{primalSDP_one_way}),
one can parametrise $\rho_{AB}$ and $\rho_{ABA^{\prime}}$ such
that some constraints are automatically fulfilled.

In particular, one can choose an operator basis of Hermitian
matrices $\{\sigma_0, \ldots, \sigma_{d^2-1}\}$ for each Hilbert
space $\mathcal{H}_d$. These matrices $\sigma_i$ can be taken
such as they satisfy the following two conditions:
$\textrm{Tr}(\sigma_i)=d\ \delta_{0i}$, and $\textrm{Tr}(\sigma_i
\sigma_j)=d\ \delta_{ij}$. In the case of qubit systems, the
Pauli matrices $\{\sigma_0, \sigma_x, \sigma_y, \sigma_z\}$ can
be selected, where the matrix $\sigma_0$ denotes the identity
operator $\openone$. For systems on $\mathcal{H}_3$, we can use
the Gell-Mann operators, that we shall denote as
$\{\sigma_i\}_{i=0}^{8}$.
With this representation, a general state
$\rho_{AB}\in\mathcal{H}^A_2\otimes\mathcal{H}^B_3$ can be
written as
\begin{equation}\label{eqnew}
  \rho_{AB}=\frac{1}{6}\mathop{\sum_{k=\{0,x,y,z\}}}_{l=0,\ldots,8} x_{kl} S_{kl},
\end{equation}
where the operators $S_{kl}=\sigma^A_k\otimes \sigma^B_l$, the
coefficients $x_{kl}$ are given by $x_{kl}=\textrm{Tr}(S_{kl}
\rho_{AB})$, and $x_{00}=\textrm{Tr}(\rho_{AB})=1$ because of
normalisation. Eq.~(\ref{eqnew}) allows us to describe any
bipartite density operator in terms of a fixed number of real
parameters $x_{kl}$.

The knowledge of Alice and Bob's POVMs $\{A_i\}$ and $\{ B_j\}$,
respectively, together with the observed probability distribution
$p_{ij}$, determines the equivalence class of compatible states
$\mathcal{S}$. Each POVM element $A_i$ and $B_j$ can also be
expanded in the appropriate operator basis as $A_i =
\sum_{k=\{0,x,y,z\}} a_{ik} \sigma_k^A$, and $B_{j} =
\sum_{l=0,\ldots,8} b_{jl} \sigma_l^B$, for some coefficients
$a_{ik}$ and $b_{jl}$, respectively. According to
Eq.~(\ref{eq_class}), to guarantee that $\rho_{AB}\in
\mathcal{S}$ [first constraint in Eq.~(\ref{primalSDP_two_way})
and in Eq.~(\ref{primalSDP_one_way})], we obtain that the
coefficients $x_{kl}$ must satisfy the following conditions,
\begin{equation}\label{cond_coeff} \sum_{kl} a_{ik} b_{jl}
x_{kl} = p_{ij}\ \ \forall i,j.
\end{equation}
That is, some coefficients $x_{kl}$ are fixed by the known
parameters $a_{ik}$, $b_{jl}$, and $p_{ij}$. Any operator
$\rho_{AB}\in \mathcal{S}$ can then always be written in the
following way,
\begin{equation}
  \label{rho_class}
  \rho_{AB}({\bf{x}})= \rho_{\text{fix}}+\sum_{kl \not\in{}I} x_{kl} S_{kl},
\end{equation}
where $\rho_{\text{fix}}$ corresponds to the part of
$\rho_{AB}({\bf{x}})$ that is completely determined by the
parameters $a_{ik}$, $b_{jl}$, and $p_{ij}$. It can be expressed
as \cite{tobi_note}
\begin{equation}
\label{rho_fix}
  \rho_{\text{fix}}=\sum_{kl \in I} x_{kl} S_{kl},
\end{equation}
where $I$ denotes a multi-index set labeling those combinations of
the indexes $k=\{0,x,y,z\}$ and $l=\{0,...,8\}$ such that
$x_{kl}$ is fixed by Eq.~(\ref{cond_coeff}). (See also
Ref.~\cite{metodo_alternativo}.)

With this representation for $\rho_{AB}({\bf x})$, the SDP given
by Eq.~(\ref{primalSDP_two_way}) can now be written as
\cite{aclar_LMI}
\begin{eqnarray}\label{primalSDP_two_way_b}
\text{minimise} && 0 \\
\nonumber \text{subject to} && \rho_{AB}({\bf
x})\oplus{}\rho_{AB}^{\Gamma}({\bf x})\geq{}0,
\end{eqnarray}
where the symbol $\oplus$ denotes direct sum. Let us compare the
second part of Eq.~(\ref{primalSDP_marcos}) with the second part
of Eq.~(\ref{primalSDP_two_way_b}).
The objective variables $x_i$ are now given by the coefficients
$x_{kl}$ of $\rho_{AB}({\bf x})$, with $kl \not\in{}I$, the
matrix $F_0$ is given by
$\rho_{\text{fix}}\oplus{}\rho_{\text{fix}}^{\Gamma}$, and the
matrices $F_i$ are those operators $S_{kl}\oplus{}S_{kl}^{\Gamma}$
with $kl \not\in{}I$.

In the SDP given by Eq.~(\ref{primalSDP_one_way}) we need to
parametrise as well the quantum state $\rho_{ABA'}$. The second
constraint in Eq.~(\ref{primalSDP_one_way}) imposes that
$\rho_{ABA'}$ must remain invariant under permutation of systems
$A$ and $A'$. This can be done with the following parametrisation
\cite{do1,doherty04,doherty05}:
\begin{eqnarray}\label{last_ap}
\rho_{ABA'}&=&\frac{1}{12} \sum_{\substack{l \\ k>m}}
f_{klm}\ (\sigma_k^A\otimes\sigma_l^B\otimes\sigma_m^{A'}\\
&+& \sigma_m^A\otimes\sigma_l^B\otimes\sigma_k^{A'}) + \sum_{kl}
f_{klk}\ \sigma_k^A\otimes\sigma_l^B\otimes\sigma_k^{A'},\nonumber
\end{eqnarray}
with $k,m=\{0,x,y,z\}$ and $l=0,\ldots,8$.

To guarantee that $\textrm{Tr}_{A'} (\rho_{ABA'}) =\rho_{AB}$
[third constraint in Eq.~(\ref{primalSDP_one_way})], the state
coefficients of $\rho_{AB}$ and $\rho_{ABA'}$ need to fulfill the
following conditions,
\begin{equation}\label{last_ap2}
f_{kl0}=x_{kl}\ \forall k,l.
\end{equation}
That is, some of the state parameters of $\rho_{ABA'}$ are already
fixed by the coefficients of $\rho_{AB}$.

To simplify the notation used later on, we shall collect the
objective variables of the SDP given by
Eq.~(\ref{primalSDP_one_way}) within two different groups of
them: The vector $\bf x$ contains those coefficients $x_{kl}$ of
$\rho_{AB}$ not fixed by Eq.~(\ref{cond_coeff}), and the vector
$\bf y$ contains those coefficients $f_{klm}$ of $\rho_{ABA'}$
not fixed by Eq.~(\ref{last_ap2}). With this parametrisation, the
first three constraints in Eq.~(\ref{primalSDP_one_way}) are
fulfilled automatically and the SDP given by
Eq.~(\ref{primalSDP_one_way}) can be reduced to solve the
following one
\begin{eqnarray}\label{primalSDP_one_way_b}
\text{minimise} && 0 \\
\nonumber \text{subject to} &&
\rho_{ABA^{\prime}}({\bf{x,y}})\geq{}0.
\end{eqnarray}

\section{Witness operators for two-way and one-way QKD}\label{sec2a}

In this section we show how to rephrase the QKD verification
criteria introduced in the previous section into a search for
appropriate witness operators. In order to do this, we use the
dual problems associated with the primal SDPs given by
Eq.~(\ref{primalSDP_two_way_b}) and
Eq.~(\ref{primalSDP_one_way_b}), respectively. In particular, we
prove that the solutions to these dual problems correspond to the
evaluation of an optimal witness operator, that belongs to the
minimal verification set of them for the given two-way or one-way
QKD protocol, on the observed data $p_{ij}$. A positive
expectation value of this optimal witness operator states that no
secret key can be distilled from the observed data $p_{ij}$.
This approach has already been considered for the symmetric
extension case in Ref.~\cite{doherty04}, and also for a slightly
different scenario in Ref.~\cite{hyllus06a}. Our main motivation
here is to show specifically that this relationship still holds
even if we restrict ourselves to partial information about the
quantum state. A detailed discussion on some duality properties
that guarantee that the solution to these dual problems can
actually be associated with a witness operator is included in
Appendix~\ref{ap_b}.

Let us first introduce the dual problem associated to the primal
SDP given by Eq.~(\ref{primalSDP_marcos}). It has the following
form \cite{vandenberghe:1996,vandenberghebook}:
\begin{eqnarray}\label{dual_marcos}
  \text{maximise}  && -\text{Tr}(F_0 Z) \\
  \nonumber
  \text{subject to} && Z\geq 0 \\
  \nonumber
              && \text{Tr}(F_i Z)=c_i\;\forall i,
\end{eqnarray}
where the Hermitian matrix $Z$ is now the objective variable.
This matrix is positive semidefinite $Z\geq 0$ and is subjected
to several linear constraints of the form $\text{Tr}(Z F_i)=c_i\;
\forall i$.

\subsection{Two-way QKD}

In this section we show that the solution to the dual problem
associated with the SDP given by Eq.~(\ref{primalSDP_two_way_b})
corresponds to the evaluation of an optimal {\it decomposable} EWs
(DEWs) \cite{optew,woro76} on the observed data $p_{ij}$. (See
also Ref.~\cite{moroder05_thesis}.) An EW $W$ is called
decomposable if and only if there exist two positive operators
$P,Q\geq{}0$, and a real parameter $\epsilon\in[0,1]$, such that
$W=\epsilon{}P+(1-\epsilon)Q^\Gamma$ \cite{optew,woro76}. In
$\mathcal{H}^A_2\otimes\mathcal{H}^B_3$ all EWs are DEWs. In what
follows, we establish this connection explicitly via the dual
problem.

The SDP given by Eq.~(\ref{primalSDP_two_way_b}) can be
transformed into a slightly different, but completely equivalent,
form as follows (see Appendix~\ref{ap_b}),
\begin{eqnarray}
  \label{PTprimal}
  \text{minimise}  && t \\
  \nonumber
  \text{subject to} && \rho_{AB}({\bf{x}}) \oplus \rho^{\Gamma}_{AB}({\bf{x}})
  + t \openone \geq 0,
\end{eqnarray}
where $t$ denotes an auxiliary objective variable. According to
Eq.~(\ref{dual_marcos}), the dual problem associated with
Eq.~(\ref{PTprimal}) can be written as
\begin{eqnarray}
  \label{PTdual}
  \text{maximise}  && -\text{Tr}[(\rho_{\text{fix}}\oplus
  \nonumber
  \rho^{\Gamma}_{\text{fix}}) Z] \\
  \nonumber
  \text{subject to} && Z \geq 0 \\
  \nonumber
              && \text{Tr}(Z)= 1 \\
              && \text{Tr}[(S_{kl} \oplus S_{kl}^{\Gamma}) Z]=0\;
              \forall kl \not \in I.
\end{eqnarray}




The structure of all the matrices which appear in this dual
problem is the direct sum of two different matrices. Then,
without loss of generality, we can assume that the same block
structure is satisfied for $Z$, {\it i.e.}, $Z=Z_1 \oplus Z_2$.
This means, in particular, that the objective function in
Eq.~(\ref{PTdual}) can now be re-expressed as
\begin{eqnarray}
  \nonumber
  \text{Tr}[(\rho_{\text{fix}}\oplus \rho^{\Gamma}_{\text{fix}}) (Z_1
  \oplus Z_2)] &=&
  \text{Tr}[(Z_1 + Z_2^{\Gamma})\rho_{\text{fix}}]\\
  \label{decomp_wit}
  &\equiv& \text{Tr}(W\rho_{\text{fix}}),
\end{eqnarray}
where we have used the property $\text{Tr}(Z_2
\rho_{\text{fix}}^{\Gamma})=\text{Tr}(Z_2^\Gamma
\rho_{\text{fix}})$ and, in the last equality, we defined the
operator $W\equiv Z_1+Z_2^{\Gamma}$. Next we show that $W$ is a
DEWs. For that, note that the semidefinite constraint $Z \geq 0$
implies $Z_1,Z_2 \geq 0$. Moreover, the witness is normalised,
since $\text{Tr}(Z)= 1$ implies $\text{Tr}(W)=\text{Tr}(Z_1 +
Z_2^{\Gamma})=1$.

To conclude, we use the remaining equality constraints,
$\text{Tr}[(S_{kl} \oplus S_{kl}^{\Gamma}) Z]=0\ \forall kl \not
\in I$, to show that to evaluate the expectation value of $W$ one
only needs to consider $\rho_{\text{fix}}$. That is, $W$ belongs
to the minimal verification set of EWs for the given QKD
protocol, and its expectation value can be obtained from the
observed data $p_{ij}$ only \cite{curty04suba}. Using the ansatz
$Z_1=\sum_{kl} z^1_{kl} S_{kl}$, and $Z_2^{\Gamma}=\sum_{kl}
z_{kl}^2 S_{kl}$, the equality constraints impose
$z_{kl}^1+z_{kl}^2=0\;\forall kl \not \in I$. Hence, the DEW $W$
has the following structure
\begin{equation}
  \label{wit_structure}
  W=\sum_{kl \in I} (z_{kl}^1 + z_{kl}^2) S_{kl} \equiv \sum_{kl \in
    I} w_{kl} S_{kl},
\end{equation}
with $w_{kl}=z_{kl}^1 + z_{kl}^2$. Combining Eq.~(\ref{rho_class})
and Eq.~(\ref{wit_structure}), we obtain
$\text{Tr}[W\rho_{AB}({\bf{x}})]=\text{Tr}(W
\rho_{\text{fix}})=\sum_{kl \in I} w_{kl} x_{kl}$.

Whenever the solution to the dual problem given by
Eq.~(\ref{PTdual}) delivers
$\text{Tr}[(\rho_{\text{fix}}\oplus\rho^{\Gamma}_{\text{fix}})
Z]\equiv{}\text{Tr}(W \rho_{\text{fix}})\geq{}0$ then no secret
key can be distilled from the observed data $p_{ij}$ with two-way
classical communication. To see this, note that, by definition,
Eq.~(\ref{PTdual}) guarantees that there exists no other DEW $W'$,
that belongs to a verification set of them for the given QKD
protocol, such that $\text{Tr}(W' \rho_{\text{fix}})<\text{Tr}(W
\rho_{\text{fix}})$.

\subsection{One-way QKD}

In this part we use the dual problem associated with the SDP
given by Eq.~(\ref{primalSDP_one_way_b}) to show that its solution
corresponds to the evaluation of an optimal witness operator for
the case of states with symmetric extensions. We shall follow the
method introduced in Ref.~\cite{doherty04}, but now we will
consider specifically the case of partial knowledge about the
quantum state. (See also Ref.~\cite{moroder05_thesis}.)

Like in the previous section, the feasibility problem given by
Eq.~(\ref{primalSDP_one_way_b}) can be transformed as follows
(see Appendix~\ref{ap_b}),
\begin{eqnarray}
  \label{primalEXT}
  \text{minimise}  && t \\
  \text{subject to} && \rho_{ABA^{\prime}}({\bf{x,y}}) + t
  \openone/d_A  \geq 0,\nonumber
\end{eqnarray}
with $d_A=\text{dim}(\mathcal{H}^{A})$, {\it e.g.}, in our case
$d_A=2$. The inclusion of the factor $d_A$ in
Eq.~(\ref{primalEXT}) does not alter its result and, as we will
see at the end of this section, it gives the correct normalisation
for the witnesses.

For convenience, we will express the state
$\rho_{ABA^{\prime}}({\bf{x,y}})$ in terms of a map $\Lambda:
\mathcal{H}^{A}_{d_A}\otimes \mathcal{H}^{B}_{d_B} \to
\mathcal{H}^{A}_{d_A}\otimes\mathcal{H}^{B}_{d_B}\otimes\mathcal{H}^{A'}_{d_{A}}$
that takes an arbitrary Hermitian operator $A=1/(d_{A}d_{B})
\sum_{kl} a_{kl} S_{kl}\in \mathcal{H}^{A}_{d_A}\otimes
\mathcal{H}^{B}_{d_B}$, with $d_B=\text{dim}(\mathcal{H}^{B})$,
to the Hermitian operator
\begin{eqnarray}\label{map}
  \Lambda(A)&=&\frac{1}{d_{A}^2d_{B}} \bigg[
\mathop{\sum_{l=\{0,...,d_{B}^2-1\}}}_{k=\{1,...,d_{A}^2-1\}}
  a_{kl}\
  (\sigma^A_k\otimes\sigma^B_l\otimes\openone^{A^{\prime}}\nonumber \\
  &+& \openone^A\otimes\sigma^B_l\otimes\sigma^{A^{\prime}}_k)\nonumber \\
  &+& \sum_{l=\{0,...,d_{B}^2-1\}}
  a_{0l}\
  \openone^A\otimes\sigma_l^B\otimes\openone^{A^{\prime}}
  \bigg],
\end{eqnarray}
Let $\rho_{\text{fix}}$ be again the part of $\rho_{AB} \in
\mathcal{S}$ that is fixed by the parameters $a_{ik}$, $b_{jl}$,
and $p_{ij}$. Without loss of generality, we consider the
following structure for $\rho_{\text{fix}}$:
$\rho_{\text{fix}}=1/(d_{A}d_{B})\sum_{kl \in I} x_{kl} S_{kl}$,
where the multi-index $I$ has the same meaning as before, {\it
i.e.}, it labels those combinations of the indexes
$k=\{0,...,d_{A}^2-1\}$ and $l=\{0,...,d_{B}^2-1\}$ such that
$x_{kl}$ is fixed by Eq.~(\ref{cond_coeff}).

Using Eq.~(\ref{map}), we can rewrite
$\rho_{ABA^{\prime}}({\bf{x,y}})$ in the following compact way,
\begin{eqnarray}
  \nonumber
  \rho_{ABA^{\prime}}({\bf{x,y}})&=& \Lambda(\rho_{\text{fix}}) +
  \sum_{kl \not \in I} x_{kl}\ \Lambda(S_{kl})\\
  \label{rho_ABA}
  &+& \sum_J y_J\ G_J,
\end{eqnarray}
where the Hermitian matrices $G_J$ can be grouped into two
different sets,
\begin{eqnarray}
  G_{klk}&=&\sigma^A_k\otimes\sigma^B_l\otimes\sigma^{A^{\prime}}_k\quad \forall l, \forall k\geq 1 \\
  G_{mlk}&=&
  \sigma^A_m\otimes\sigma^B_l\otimes\sigma^{A^{\prime}}_k\nonumber
  \\
  &+&\sigma^A_k\otimes\sigma^B_l\otimes\sigma^{A^{\prime}}_m \quad \forall l, \forall k>m\geq 1,\nonumber
\end{eqnarray}
and where the multi-index $J$ is used to label both different
combinations of the indices $k, l$, and $m$.

The dual problem associated with Eq.~(\ref{primalEXT}) can now be
written as
\begin{eqnarray}
  \label{dualEXT}
  \text{maximise}  && -\text{Tr}[Z \Lambda(\rho_{\text{fix}})] \\
  \nonumber
  \text{subject to} && Z \geq 0 \\
  \nonumber
              && \text{Tr}(Z)=d_A \\
  \nonumber
              && \text{Tr}[Z \Lambda(S_{kl})]= 0\ \;\forall
              kl \not \in I \\
  \nonumber
             && \text{Tr}(Z G_J)=0\ \;\forall J.
\end{eqnarray}

Next we search for the most general form of a possible solution
$Z$ for this dual problem. It will enable us to extract the most
compact form of a witness operator for the symmetric extendibility
problem.

All the linear constraints on the operator $Z$ contained in
Eq.~(\ref{dualEXT}), as well as the objective function itself,
are invariant under the swap operator $P$, which exchanges the
first and the third subsystem. Moreover, the positive
semidefinite constraint $\bar Z= P Z P \geq 0$ is also satisfied
since $P$ is a unitary operator, {\it i.e.}, $P^2=\openone$. This
means that, if $Z$ is a solution for the dual problem, the
operator $\bar Z$ is also a possible solution for it, since it
fulfills all the constraints and it gives exactly the same
expectation value. Following a similar argumentation, also the
equal mixture of $Z$ and $\bar Z$, {\it i.e.}, $\tilde Z=1/2(Z +
\bar Z )$, is as well a possible solution. Therefore, without
loss of generality, we can consider that $Z$ is invariant under
the swap operator $P$. Under this assumption, it turns out that
$Z$ can be decomposed as follows
\begin{eqnarray}\label{new_m}
Z&=&\frac{1}{d_{A}^2d_{B}}\bigg[ \mathop{\sum_{l}}_{k>m} z_{mlk}\
(\sigma^A_m
\otimes\sigma^B_l\otimes\sigma^{A^{\prime}}_k \\
&+& \sigma^A_k\otimes\sigma^B_l\otimes \sigma^{A^{\prime}}_m) +
\sum_{kl} z_{klk}\
\sigma^A_k\otimes^B_l\otimes\sigma^{A^{\prime}}_k\bigg].\nonumber
\end{eqnarray}
Let us now analyse in more detail the linear constraints on $Z$
given in Eq.~(\ref{primalEXT}). Each linear constraint cancels
one of the coefficients $z_{mlk}$. For instance, the constraint
$\text{Tr}(ZG_{iji})$ imposes $z_{iji}=0$. Then, we can remove all
these linear constraints, except the normalisation condition
$\text{Tr}(Z)=d_A$, by just making the proper coefficients
$z_{mlk}$ in Eq.~(\ref{new_m}) equal to zero. This way we arrive
at the following form for the variable $Z$, which we shall denote
by $Z^*$,
\begin{eqnarray}\label{formZ2}
  Z^*&=&\frac{1}{d_{A}^2d_{B}} \bigg[ \mathop{\sum_{l;k\geq 1}}_{kl \in I}
  z_{kl1}(\sigma^A_k\otimes\sigma^B_l\otimes\openone^{A^{\prime}} \\
  &+& \openone^A\otimes\sigma^B_l\otimes\sigma^{A^{\prime}}_k)
  + \mathop{\sum_{l}}_{0l \in I} z_{0l0}\
  \openone^A\otimes\sigma^B_l
  \otimes\openone^{A^{\prime}}\bigg].\nonumber
\end{eqnarray}
That is, if we assume the form $Z^*$ for the variable $Z$ in the
dual problem given by Eq.~(\ref{dualEXT}), then the linear
constraints are fulfilled automatically. Substituting the
variable $Z$ with $Z^*$ in Eq.~(\ref{dualEXT}) we obtain the
following shorter form for the dual problem,
\begin{eqnarray}
  \label{dualEXT2}
  \text{maximise}  && -\text{Tr}[Z^* \Lambda(\rho_{\text{fix}})] \\
  \nonumber
  \text{subject to} && Z^* \geq 0 \\
  \nonumber
              && \text{Tr}(Z^*)=d_A.
\end{eqnarray}
Now, in order to extract a witness operator from the Hermitian
operator $Z^*$, we follow the method proposed in
Ref.~\cite{doherty04}. In particular, every map $\Lambda:
\mathcal{H}^{A}_{d_A}\otimes \mathcal{H}^{B}_{d_B} \to
\mathcal{H}^{A}_{d_A}\otimes\mathcal{H}^{B}_{d_B}\otimes\mathcal{H}^{A'}_{d_{A}}$
has associated an adjoint map $\Lambda^{\dag}:
\mathcal{H}^{A}_{d_A}\otimes\mathcal{H}^{B}_{d_B}\otimes\mathcal{H}^{A'}_{d_{A}}
\to \mathcal{H}^{A}_{d_A}\otimes \mathcal{H}^{B}_{d_B}$ defined as
$\text{Tr}[U\ \Lambda(V)]=\text{Tr}[\Lambda^{\dag}(U)\ V]$ for any
Hermitian operators $V\in \mathcal{H}^{A}_{d_A}\otimes
\mathcal{H}^{B}_{d_B}$, and $U \in
\mathcal{H}^{A}_{d_A}\otimes\mathcal{H}^{B}_{d_B}\otimes\mathcal{H}^{A'}_{d_{A}}$.
With this definition, we can rewrite the objective function in
Eq.~(\ref{dualEXT2}) as
\begin{equation}
  \text{Tr}[Z^*\Lambda(\rho_{\text{fix}})] =
  \text{Tr}[\Lambda^{\dag}(Z^*)\rho_{\text{fix}}] \equiv
  \text{Tr}(W_{\text{sym}} \rho_{\text{fix}}),
\end{equation}
where we defined $W_{\text{sym}}\equiv \Lambda^{\dag}(Z^*)$ as the
desired witness operator for the symmetric extendibility problem.
In the remaining part of this section, we obtain the general
structure of the witness operator $W_{\text{sym}}$. For that, we
simply apply the adjoint map $\Lambda^{\dag}$ to the operator
$Z^*$, set the resulting operator equal to an operator of
arbitrary form $W^*$, solve the equality constraint, and,
finally, formulate the dual problem in terms of this new operator
$W^*$.

The map $\Lambda^{\dag}$ can be written as \cite{doherty04},
\begin{equation}
  \Lambda^{\dag}(Z)=\frac{1}{d_A}
  \big[\text{Tr}_{A'}(Z)+\text{Tr}_{A^{\prime}}(P Z P) - \frac{1}{d_A}
  \openone^A\otimes\text{Tr}_{AA^{\prime}}(Z)\big].
\end{equation}
Setting $\Lambda^{\dag}(Z^*)$ equal to an arbitrary Hermitian
operator $W^*=1/(d_{A}d_{B}) \sum_{kl} w_{kl} S_{kl}$ we obtain
the following equality constraint
\begin{equation}
  \sum_{kl} w_{kl} S_{kl}=
  \mathop{\sum_{l;k\geq 1}}_{kl \in I} \frac{2z_{kl0}}{d_A}
  S_{kl}
  + \mathop{\sum_{l}}_{0l \in I} z_{0l0}\
  S_{0l}.
\end{equation}
Since we have expressed every operator in terms of an operator
basis, the equality constraint can only be fulfilled if the
coefficients $z_{klm}$ of $Z^*$, and the coefficients $w_{kl}$ of
$W^*$, are related via:
\begin{eqnarray}
  \label{connectionZWsym}
  w_{kl}&=&2  z_{kl0}/d_A\;\;\;\;\forall l, \forall k\geq 1, kl \in I, \\
  w_{0l}&=& z_{0l0}/d_A\;\;\;\;\;\; \forall l, 0l \in I,\nonumber \\
  w_{kl}&=&0\;\;\;\;\;\;\;\;\;\;\;\;\;\;\;\; kl \not\in
  I.\nonumber
\end{eqnarray}
Now, instead of considering the matrix $Z^*$ as the objective
variable of the dual problem, we can equivalently consider the
matrix $W^*$ as the free variable. In order to do so, we only need
to translate the positive semidefinite constraint $Z^*\geq 0$
together with the normalisation condition $\text{Tr}(Z^*)=d_A$
included in Eq.~(\ref{dualEXT2}) into new constraints on $W^*$.
This can be done by using Eq.~(\ref{connectionZWsym}). This way,
we arrive at the following form for the dual problem:
\begin{eqnarray}
  \label{dualEXT3}
  \text{maximise}  && -\text{Tr}(W^* \rho_{\text{fix}}) \\
  \nonumber
  \text{subject to} && W^*\otimes\openone^{A^{\prime}} +
  P(W^*\otimes\openone^{A^{\prime}})P\geq 0 \\
  \nonumber   && \text{Tr}(W^*)=1,
\end{eqnarray}
where the variable $W^*$ represents a witness operator for the
symmetric extendibility problem. Moreover, from
Eq.~(\ref{connectionZWsym}) we obtain that $W^*$ can always be
expressed as
\begin{equation}
  W^*=\frac{1}{d_{A}d_{B}}\sum_{kl \in I} w_{kl} S_{kl}.
\end{equation}
That is, $W^*$ belongs to the minimal verification set of
witnesses for the given one-way (RR) QKD protocol. Like in the
previous section, whenever the solution to the dual problem given
by Eq.~(\ref{dualEXT3}) delivers $\text{Tr}(W^*
\rho_{\text{fix}})\geq{}0$ then no secret key can be distilled
from the observed data $p_{ij}$ with one-way RR. The case of
one-way QKD with DR can be analysed in a similar way.

\section{Evaluation}\label{section_example}

In this section we study the two-state QKD protocol \cite{ben92},
both for the case of two-way and one-way classical communication.
The analysis for other qubit-based QKD schemes is completely
analogous, and we include very briefly the results of our
investigations on other well-known QKD protocols in
Appendix~\ref{other_QKD}. We refer here to single-photon
implementations of the qubit. The state of the qubit is
described, for instance, by some degree of freedom in the
polarisation of the photon. In our calculations we follow the
approach introduced in Sec.~\ref{verif_criteria}, although
similar results could also be obtained using the witness approach
presented in Sec.~\ref{sec2a}. The numerical evaluations are
performed with the freely-available SDP solver SDPT3-3.02
\cite{sdpt}, together with the input tool YALMIP \cite{yalmip}.


We shall consider that the observed joint probability distribution
$p_{ij}$ originates from Alice and Bob measuring the following
quantum state
\begin{eqnarray}\label{channel_eq}
\rho_{AB}&=&(1-p)\bigg[(1-e)\openone^A\otimes{}U^B(\theta)
\ket{\psi}_{AB}\bra{\psi}\openone^A\otimes{}U^{B\dag}(\theta)\nonumber\\
&+&\frac{e}{2}\rho_A\otimes\tilde{\openone}^B\bigg]+p\rho_A\otimes\ket{vac}_B\bra{vac},
\end{eqnarray}
where $p\in[0,1]$ denotes the probability that Bob receives the
vacuum state $\ket{vac}_B$, $e\in[0,1]$ represents an error
parameter (or depolarising rate) of the channel, $\openone^A$ is
the identity operator on Alice's Hilbert space, $U^B(\theta)$
represents a unitary operator acting on Bob's system,
$\ket{\psi}_{AB}$ denotes the effective bipartite state initially
prepared by Alice in the given QKD protocol, $\rho_A$ represents
Alice's reduced density matrix ({\it i.e.},
$\rho_A=Tr_B(\ket{\psi}_{AB}\bra{\psi})$), and the operator
$\tilde{\openone}^B$ is given by
$\tilde{\openone}^B=\openone^B-\ket{vac}_B\bra{vac}$.

The quantum state given by Eq.~(\ref{channel_eq}) defines one
possible eavesdropping interaction. But our analysis can
straightforwardly be applied to other quantum channels, as it
depends only on the probability distribution $p_{ij}$ that
characterises the results of Alice's and Bob's measurements. We
include the operator $U^B(\theta)$ in Eq.~(\ref{channel_eq}) to
model the collective noise (or correlated noise) introduced by
the quantum channel ({\it e.g.}, optical fiber)
\cite{yamamoto,boi}. This noise arises from the fluctuation of
the birefringence of the optical fiber which alters the
polarisation state of the photons. When this fluctuation is slow
in time, its effect can be
described with a unitary operation \cite{yamamoto,boi}. For
simplicity, we shall consider that $U^B(\theta)$ is parametrised
only with one real parameter $\theta$. In particular, we choose
$U^B(\theta)=\cos\theta\ket{0}\bra{0}-\sin\theta\ket{0}\bra{1}+
\sin\theta\ket{1}\bra{0}+\cos\theta\ket{1}\bra{1}+\ket{vac}\bra{vac}$
with $\theta\in[0,\pi/4]$. If $\theta=0$ no collective noise is
present and Eq.~(\ref{channel_eq}) describes a depolarising
channel with loss.

In order to illustrate our results, we calculate an upper bound on
the tolerable depolarising rate $e$ as a function of the photon
loss probability $p\in[0,1]$. Moreover, for simplicity, we take
only two different values of the angle $\theta$. For instance, we
choose $\theta=0$ and $\theta=\pi/8$. These three parameters,
$e$, $p$, and $\theta$, allow us to evaluate the performance of a
QKD protocol when the quantum channel is described by
Eq.~(\ref{channel_eq}). One could also select other figures of
merit in order to evaluate a protocol, such as the quantum bit
error rate (QBER). This is the rate of events where Alice and Bob
obtain different results. It refers to the sifted key, {\it i.e},
it considers only those events where the signal preparation and
detection methods employ the same polarisation basis. We include
as well an analytic expression for the QBER for the given QKD
protocol.


\subsection{Two-state protocol}\label{sec_two-state}

The two-state protocol \cite{ben92} is one of the simplest QKD
protocols. It is based on the random transmission of only two
nonorthogonal states, $|\varphi_0\rangle$ and
$|\varphi_1\rangle$. Alice chooses, at random and independently
every time, a bit value $i$, and prepares a qubit in the state
$|\varphi_i\rangle=\alpha|0\rangle+(-1)^i\beta|1\rangle$, with
$0<\alpha<1/\sqrt{2}$ and $\beta=\sqrt{1-\alpha^2}$, that is sent
it to Bob. On the receiving side, Bob measures the qubit he
receives in a basis chosen at random within the set
$\{\{|\varphi_0\rangle,|\varphi_0^{\perp}\rangle\},\{
|\varphi_1\rangle,|\varphi_1^{\perp}\rangle\}\}$, with
$|\langle\varphi_i|\varphi_i^{\perp}\rangle|=0$. The loss of a
photon corresponds to a projection onto the vacuum state
$\ket{vac}$. Bob could also employ a different detection method
defined by a POVM with the following operators:
$B_j=1/(2\beta^2)|\varphi_{1-j}^{\perp}\rangle\langle\varphi_{1-j}^{\perp}|$
with $j=0,1$, $B_{null}=\ket{0}\bra{0}+\ket{1}\bra{1}-\sum_j B_j$,
and $B_{vac}=\ket{vac}\bra{vac}$. In this last case, Alice's bit
value $i$ is associated with the operator $B_i$, while the
operator $B_{null}$ represents an inconclusive result. This is the
approach that we shall consider here.

The preparation process can be thought of as Alice prepares first
the bipartite signal state
$|\psi\rangle_{AB}=1/\sqrt{2}(|0\rangle_A|\varphi_0\rangle_B+
|1\rangle_A|\varphi_1\rangle_B)$, and then she measures her first
subsystem with the POVM operators $A_{i}=\ket{i}\bra{i}$ with
$i=0,1$. The fact that the reduced density matrix of Alice is
fixed and cannot be modified by Eve is vital to guarantee the
security of this scheme. Otherwise, the joint probability
distribution $p_{ij}$ alone does not allow Alice and Bob to
distinguish between the entangled state $|\psi\rangle_{AB}$ and
the separable one $\sigma_{AB}=1/2\sum_{i=0}^1
|i\rangle_{A}\langle{}i|\otimes|\varphi_i\rangle_{B}\langle\varphi_i|$
\cite{curty04a,curty04suba}. We need to add then to the
observables given above also the operators
$\sigma_x\otimes\sigma_0$ and $\sigma_y\otimes\sigma_0$ such as
Alice has complete tomographic knowledge of $\rho_A$.

\begin{figure}
\begin{center}
\includegraphics[angle=-90,scale=.32]{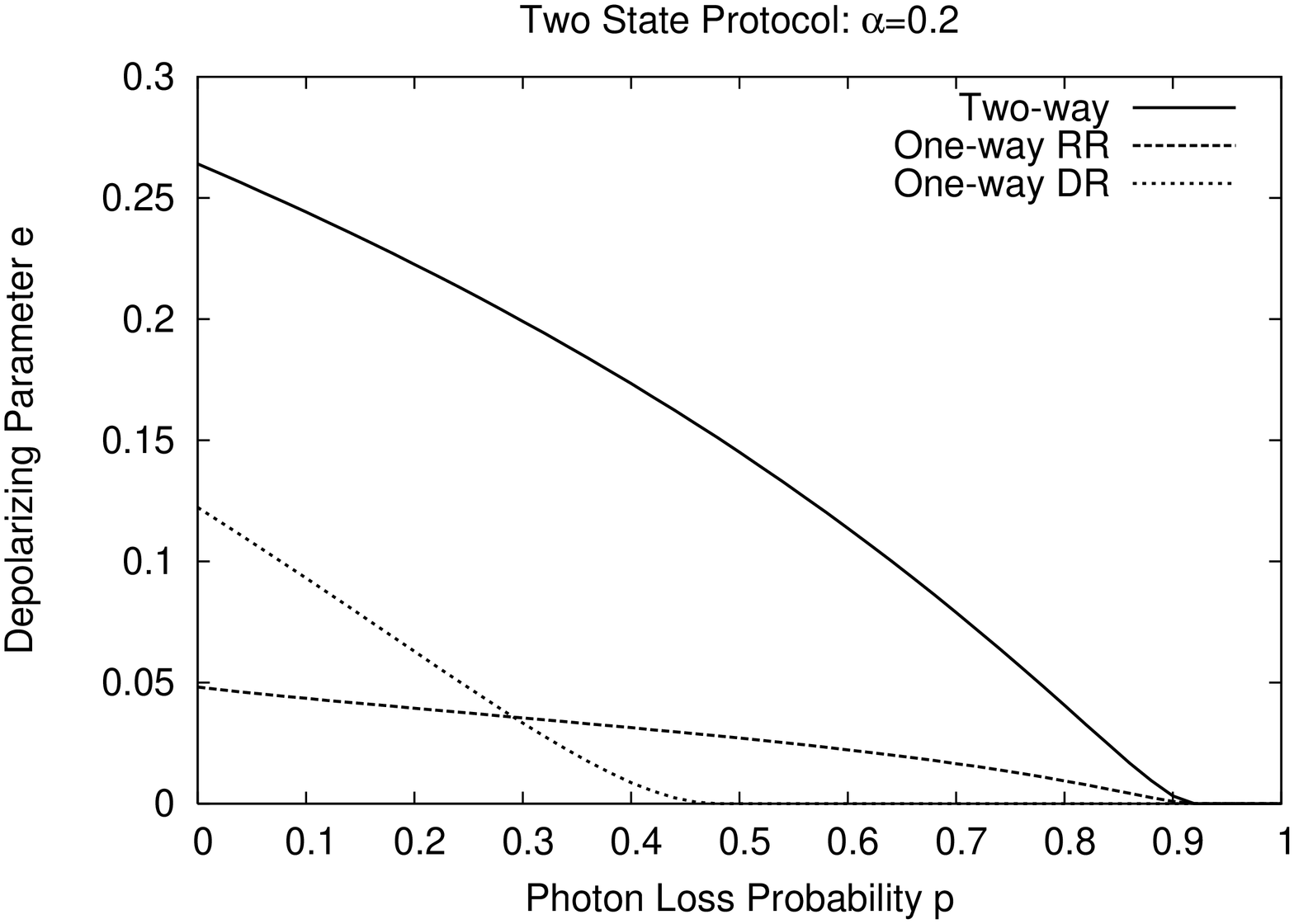}
\includegraphics[angle=-90,scale=.32]{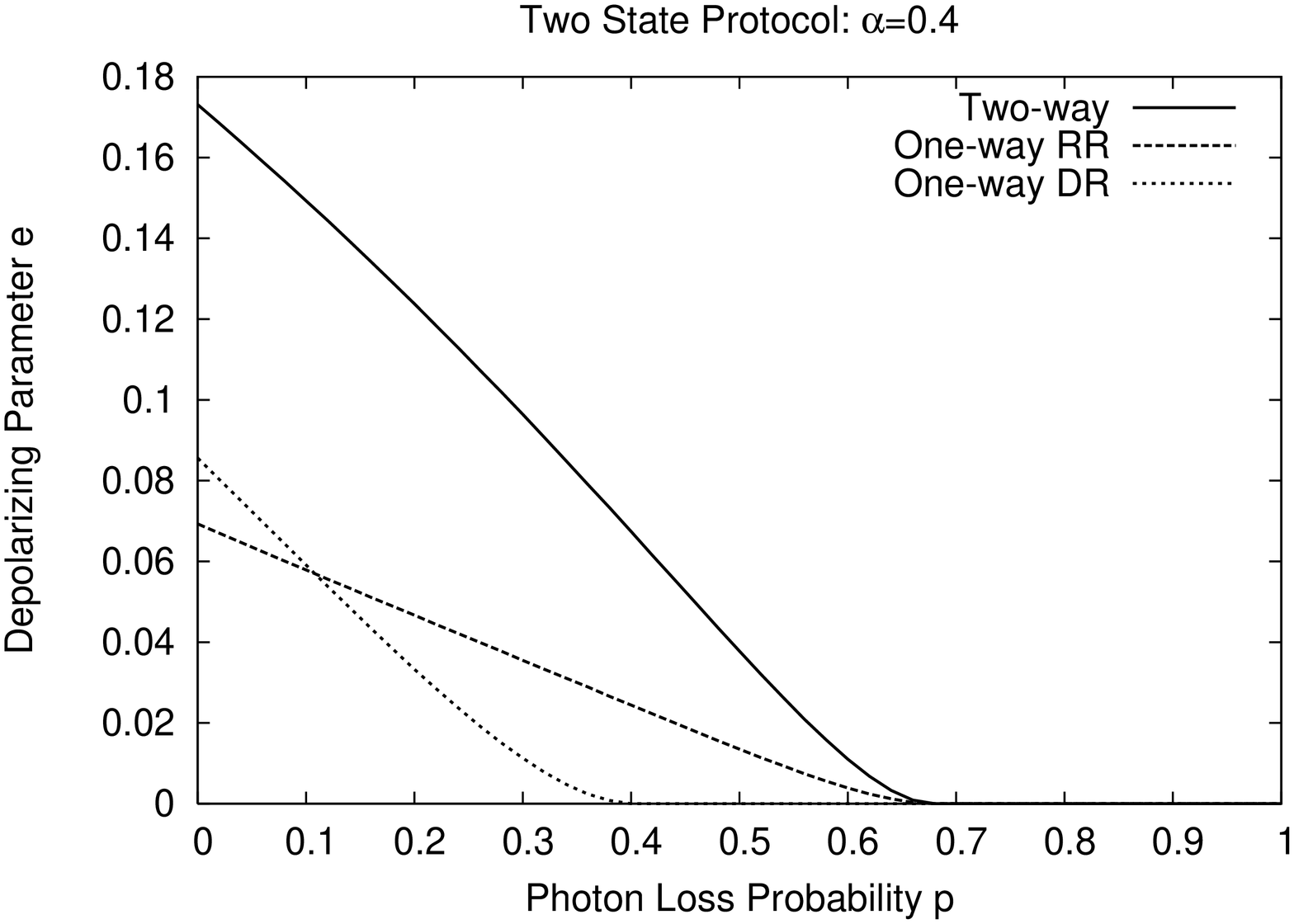}
\end{center}
\caption{Upper bound on the depolarising rate $e$ as a function of
the photon loss probability $p$ for the two-state QKD protocol.
The equivalence class of states $\mathcal{S}$ is fixed by the
observed data $p_{ij}$, which are generated via measurements onto
the state given by Eq.~(\ref{channel_eq}). Two-way classical
post-processing is illustrated with a solid line. One-way
classical post-processing is represented with a dashed line
for RR, and with a dotted line for DR. The cases $\theta=0$
and $\theta=\pi/8$ provide the same results. It states that no
secret key can be obtained from the correlations established by
the users. \label{two-state-fig}}
\end{figure}

Following the approach introduced in Sec.~\ref{verif_criteria}, in
Fig.~\ref{two-state-fig} we present an upper bound on the
tolerable depolarising rate $e$ as a function of the photon loss
probability $p$ for two different values of the parameter
$\alpha$. It states that no secret key can be distilled from the
correlations established by the users.
In this example, the results obtained coincide when $\theta=0$ and
$\theta=\pi/8$. To obtain an upper bound on the tolerable QBER
one can use the following expression
\begin{equation}\label{qber_2}
QBER=\frac{2\sin^2\theta+(1-2\sin^2\theta)e}
{2\{2\beta^2+(\alpha^2-\beta^2)[\gamma+(1-\gamma)e]\}},
\end{equation}
with $\gamma=2(\alpha^2\sin^2\theta+\beta^2\cos^2\theta)$. In
particular, for given values of the parameters $\alpha$, $\beta$,
and $\theta$, one only needs to substitute in Eq.~(\ref{qber_2})
the value of $e$ given in Fig.~\ref{two-state-fig} as a function
of the parameter $p$.

Remarkably, the cut-off point for two-way QKD presented in
Fig.~\ref{two-state-fig}, {\it i.e.}, the value of the photon
loss probability $p$ that makes $e=0$ and also $QBER=0$,
coincides with the limit imposed by the unambiguous state
discrimination attack
\cite{chefles,ivanovic,peresusd,dieks,dusek}.
In the two-state protocol this limit is given by $p=1-2\alpha^2$.
(See also Ref.~\cite{tamakiusd}.) Fig.~\ref{two-state-fig} also
shows a difference between one-way classical post-processing with
RR and with DR as a function of the parameter $p$. The reason
behind this effect is beyond the scope of this paper and needs
further investigations.

\section{CONCLUSION}

A fundamental question in quantum key distribution (QKD) is to
determine whether the legitimate users of the system can use their
available measurement results to generate a secret key via
two-way or one-way classical post-processing of the observed
data. In this paper we have investigated single-photon QKD
protocols in the presence of loss introduced by the quantum
channel. Our results are based on a simple precondition for
secure QKD for two-way and one-way classical communication. In
particular, the legitimate users need to prove that there exists
no separable state (in the case of two-way QKD), or that there
exists no quantum state having a symmetric extension (one-way
QKD), that is compatible with the available measurements results.

We have shown that both criteria can be formulated as a convex
optimisation problem known as a primal semidefinite program (SDP).
Such instances of convex optimisation problems can be solved
efficiently, for example by means of interior-point methods.
Moreover, these SDP techniques allow us to evaluate these
criteria for any single-photon QKD protocol in a completely
systematic way. A similar approach was already used in
Ref.~\cite{tobione} for the case of one-way QKD without losses.
Here we complete these results, now presenting specifically the
analysis for the case of a lossy channel. Furthermore, we have
shown that these QKD verification criteria based on SDP provide
also a means to search for witness operators for a given two-way
or one-way QKD protocol. Any SDP has an associated dual problem
that represents also a SDP. We have demonstrated that the
solution to the this dual problem corresponds to the evaluation
of an optimal witness operator that belongs to the minimal
verification set of them for the given two-way (or one-way) QKD
protocol. Most importantly, a positive expectation value of this
optimal witness operator guarantees that no secret key can be
distilled from the available measurements results. Finally, we
have illustrated our results by analysing the performance of
several well-known qubit-based QKD protocols for a given channel
model.

\section{ACKNOWLEDGEMENTS}

We gladly acknowledge stimulating discussions with K. Tamaki, O.
G\"uhne, C.-H. F. Fung, and X. Ma. We specially thank H.-K. Lo
and N. L\"utkenhaus for their critical discussion of this
article, and F. Just for help on the numerics. Financial support
from NSERC, CIPI, CRC program, CFI, OIT, CIAR, PREA, DFG under
the Emmy Noether programme, and the European Commission
(Integrated Project SECOQC) are gratefully acknowledged. M.C.
also thanks the financial support from a Post-doctoral grant from
the Spanish Ministry of Science (MEC).

\appendix




\section{Some duality properties}\label{ap_b}

In this Appendix we present some duality properties of a general
SDP \cite{vandenberghe:1996,vandenberghebook} that guarantee that
the solution to the dual problems introduced in Sec.~\ref{sec2a}
can be associated with a witness operator.

The primal problem given by Eq.~(\ref{primalSDP_marcos}) is called
feasible (strictly feasible) if there exists $\bf x$ such as
$F({\bf x})\geq 0$ ($F({\bf x})>0$). Similarly, the dual problem
given by Eq.~(\ref{dual_marcos}) is called \emph{feasible}
(\emph{strictly feasible}) if there exists a matrix $Z \geq 0$
($Z>0$) which fulfills all the desired constraints.

The {\it weak duality condition}, illustrated in
Eq.~(\ref{weakduality}), allows to derive simple upper and lower
bounds for the solution of either the primal or dual problem. In
particular, for every feasible solution ${\bf{x}}$ of the primal
problem and for every feasible solution $Z$ of the dual problem,
the following relation holds:
\begin{equation}
\label{weakduality}
  c^T{\bf{x}}+\text{Tr}(Z F_0)=\text{Tr}(Z F({\bf{x}})) \geq 0.
\end{equation}

The {\it strong duality condition} certifies whether the optimal
solution to the primal and dual problem, that we shall denote as
$p^*$ and $d^*$, respectively, are equal. More precisely,
$p^*=d^*$ if (1) The primal problem is strictly feasible, or (2)
The dual problem is strictly feasible. Moreover, if both
conditions are satisfied simultaneously then it is guaranteed that
there is a feasible pair
$({\bf{x_{\text{opt}}}},{Z_{\text{opt}}})$ achieving the optimal
values $p^*=d^*$. This last condition is known as the {\it
complementary slackness condition}.

The SDP given by Eq.~(\ref{primalSDP_marcos}), when $c=0$, can
always be transformed as follows
\cite{vandenberghe:1996,vandenberghebook}
\begin{eqnarray}
  \label{primalSDP}
  \text{minimise}  && t \\
  \nonumber
  \text{subject to} && F({\bf{x}},t)=F({\bf{x}}) + t \openone \geq 0.
\end{eqnarray}
This SDP is always strictly feasible. To see this, note that if
${\bf{x}}=0$ and $t>|\min_i \lambda_i(F_0)|$, where
$\lambda_i(F_0)$ denote the eigenvalues of the matrix $F_0$, then
$F({\bf{x}},t)>0$. Moreover, it can be shown that
Eq.~(\ref{primalSDP}) is equivalent to the original SDP. Let $t^*$
be the solution to Eq.~(\ref{primalSDP}). If $t^* > 0$, the
original problem is infeasible since $F({\bf{x}}) \not\geq
0\;\forall\ {\bf{x}}$. On the other hand, if $t^*\leq 0$ there
exists $\bar{\bf{x}}$ such that $F(\bar{\bf{x}}) \geq 0$, stating
that the original problem is feasible. That is, the solution $t^*$
of the SDP given by Eq.~(\ref{primalSDP}) certifies whether the
original problem is indeed feasible or not.

The dual problem associated with Eq.~(\ref{primalSDP}) is given
by:
\begin{eqnarray}
  \label{dualSDP}
  \text{maximise}  && -\text{Tr}(F_0 Z) \\
  \nonumber
  \text{subject to} && Z\geq 0 \\
  \nonumber
              && \text{Tr}(F_i Z)=0\;\forall i, \\
  \nonumber
              && \text{Tr}(Z)=1.
\end{eqnarray}
If all the matrices $F_i$ are traceless, {\it i.e.},
$\text{Tr}(F_i)=0$, this dual problem is also always strictly
feasible. A trivial strictly feasible solution to this problem is
given by $Z=\openone/d
> 0$, where $d$ denotes the dimension of $F_i$.

If we apply the three duality conditions mentioned above to the
SDPs given by Eq.~(\ref{primalSDP}) and Eq.~(\ref{dualSDP}) we
find that, if Eq.~(\ref{primalSDP}) delivers an infeasible
solution $t^*>0$,
\begin{equation}
\label{violation}
  \text{Tr}(F_0 Z_{\text{opt}})=-d^*=-t^* < 0.
\end{equation}
This arises from the fact that the strong duality relation
guarantees that $d^*=t^*$, and the complementary slackness
condition certifies that there is a $Z_{\text{opt}}$ that
achieves the optimal value $d^*$. When $t^* \leq 0$ the weak
duality condition assures that
\begin{equation}
\label{requirement}
  \text{Tr}(F_0 Z) \geq -t^* \geq 0
\end{equation}
for every feasible solution $Z$ of the dual problem. Both results
together show that the solution to the dual problem given by
Eq.~(\ref{dualSDP}) can be associated to a witness operator $W$.
In particular, the ability of $W$ to detect, at least, one state,
{\it i.e.}, $\exists\ \rho$ such as $\text{Tr}(W\rho) < 0$, can
be related with Eq.~(\ref{violation}). On the other hand, the
requirement that $W$ is positive on all states belonging to a
given set of them can be related with Eq.~(\ref{requirement}). To
achieve the desired equivalence, however, the dual problem must be
strictly feasible, otherwise the complementary slackness
condition does not hold and the existence of an appropriate
witness is not guaranteed. It turns out that all the linear
constraints included in the dual problems considered in
Sec.~\ref{sec2a} have traceless matrices $F_i$, such that these
dual problems are always strictly feasible.

\section{More qubit-based QKD schemes}\label{other_QKD}

In this Appendix we include very briefly the results of our
investigations on other well-known qubit-based QKD protocols.
Like in Sec.~\ref{section_example}, we shall consider that the
observed data $p_{ij}$ are generated via measurements onto the
state given by Eq.~(\ref{channel_eq}).


\subsection{Six-state protocol}

In this scheme, Alice prepares a qubit in one of the following six
quantum states: $\{\ket{0}, \ket{1},
\ket{\pm}=1/\sqrt{2}(\ket{0}\pm\ket{1}),
\ket{\tilde{\pm}}=1/\sqrt{2}(\ket{0}\pm{}i\ket{1})$, and sends it
to Bob \cite{bruss98a}. On the receiving side, Bob measures each
incoming signal by projecting it onto one of the three possible
bases. The loss of a photon in the channel is characterised by a
projection onto the vacuum state $\ket{vac}$.

\begin{figure}
\begin{center}
\includegraphics[angle=-90,scale=.32]{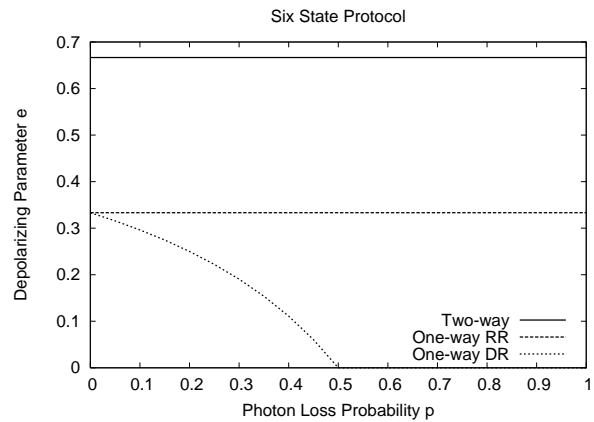}
\end{center}
\caption{Upper bound on the depolarising rate $e$ as a function of
the photon loss probability $p$ for the six-state QKD protocol.
Two-way classical post-processing is illustrated with a solid
line, while one-way classical post-processing is represented with
a dashed line for RR, and with a dotted line for DR. The cases
$\theta=0$ and $\theta=\pi/8$ provide the same results in this
case.\label{six-state-fig}}
\end{figure}

The resulting upper bound on the tolerable depolarising rate $e$
is illustrated in Fig.~\ref{six-state-fig}.
The QBER is given by
\begin{equation}
QBER=\frac{1}{6}\bigg[4\sin^2\theta+(3-4\sin^2\theta)e\bigg].
\end{equation}
For $\theta=0$ we find, as expected, that whenever
$QBER\geq{}33\%$ (corresponding to a value of $e=0.66$) no secret
key can be distilled by two-way classical post-processing
\cite{bruss98a,hut94}. In the case of one-way classical
post-processing (both for RR and DR), and assuming $\theta=0$ and
$p=0$, we obtain that secure QKD might only be possible for a
$QBER<1/6$ ($e=0.33$). A possible eavesdropping strategy to
attain this cut-off point is, for instance, to use an universal
cloning machine to clone every signal sent by Alice such as the
fidelities of Eve's and Bob's clones coincide \cite{bech}. (See
also Ref.~\cite{tobione}.)

\subsection{Four-state protocol}\label{four_sec}

The four-state protocol \cite{bennett84a} is similar to the
six-state protocol, but now Alice sends one of four possible
signal states instead of one of six. In particular, she chooses
one state within the set $\{\ket{0}, \ket{1}, \ket{\pm}\}$ and
sends it Bob. Each received signal is projected by Bob onto one
of the two possible bases, together with a projection onto the
vacuum state $\ket{vac}$ corresponding to the loss of a photon.

\begin{figure}[h]
\begin{center}
\includegraphics[angle=-90,scale=.32]{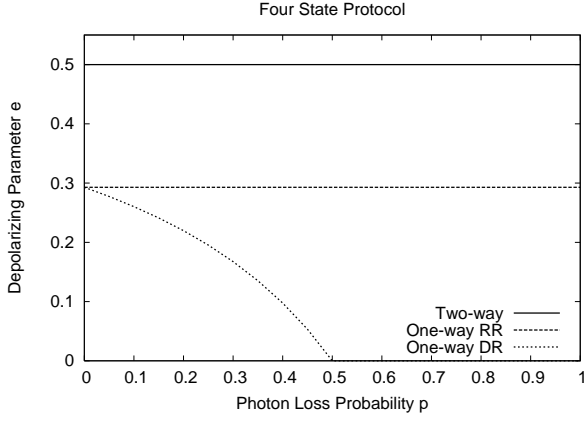}
\end{center}
\caption{Upper bound on the depolarising rate $e$ as a function of
the photon loss probability $p$ for the four-state QKD protocol.
Two-way classical post-processing is illustrated with a solid
line, while one-way classical post-processing is represented with
a dashed line for RR, and with a dotted line for DR. The cases
$\theta=0$ and $\theta=\pi/8$ provide the same results in this
case. This upper bound also coincides for the case of the trine
protocol and for the QKD scheme proposed in Ref.~\cite{ac06} when
$\theta=0$. \label{four-state-fig}}
\end{figure}

The resulting upper bound on the depolarising rate $e$ is
illustrated in Fig.~\ref{four-state-fig}.
The QBER is now given by
\begin{equation}
QBER=\sin^2\theta+\frac{(1-2\sin^2\theta)e}{2}.
\end{equation}
If $\theta=0$ we obtain the well-known result stating that
whenever $QBER\geq{}25\%$ (corresponding to a value of $e=0.5$) no
secret key can be distilled by two-way classical post-processing
\cite{hut94}. Similarly, for the case of one-way classical
post-processing, and assuming $\theta=0$ and $p=0$, we find that
the $QBER$ must be lower than $14,6\%$ ($e=0.292$). This last
result coincides with the value of the QBER produced by an
eavesdropping strategy where Eve and Bob Shannon information are
equal \cite{fuchs96,cirac97}.

\subsection{Qubit-based four-plus-two-state protocol}

This scheme can be seen as a combination of two two-state QKD
protocols \cite{hut95,notefpt}. More precisely, Alice selects, at
random and independently each time, one of the following four
signal states,
$\{|\varphi_k\rangle=\alpha|0\rangle+(-1)^k\beta|1\rangle,
|\varphi_{\bar{k}}\rangle=\alpha|0\rangle+i(-1)^k\beta|1\rangle\}$
with $k=0,1$, and sends it to Bob. On the receiving side, Bob
measures each incoming signal by choosing, at random and
independently for each signal, one of two possible POVMs. Each
POVM corresponds to the one used in the two-state protocol (see
Sec.~\ref{sec_two-state}) for the signal states
$|\varphi_k\rangle=\alpha|0\rangle+(-1)^k\beta|1\rangle$, with
$k=0,1$, and
$|\varphi_{\bar{k}}\rangle=\alpha|0\rangle+i(-1)^k\beta|1\rangle$,
with $k=0,1$, respectively.

\begin{figure}
\begin{center}
\includegraphics[angle=-90,scale=.32]{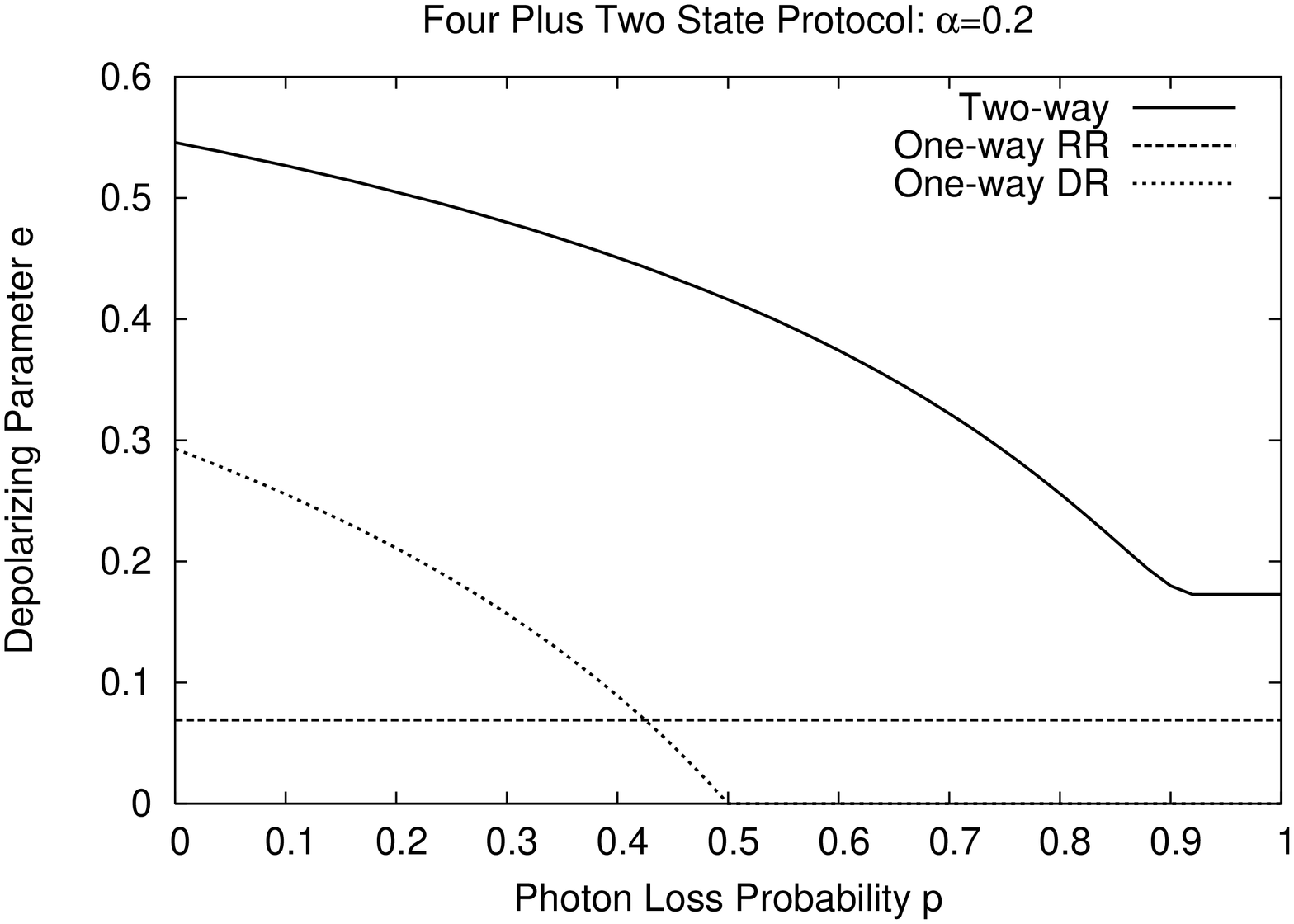}
\includegraphics[angle=-90,scale=.32]{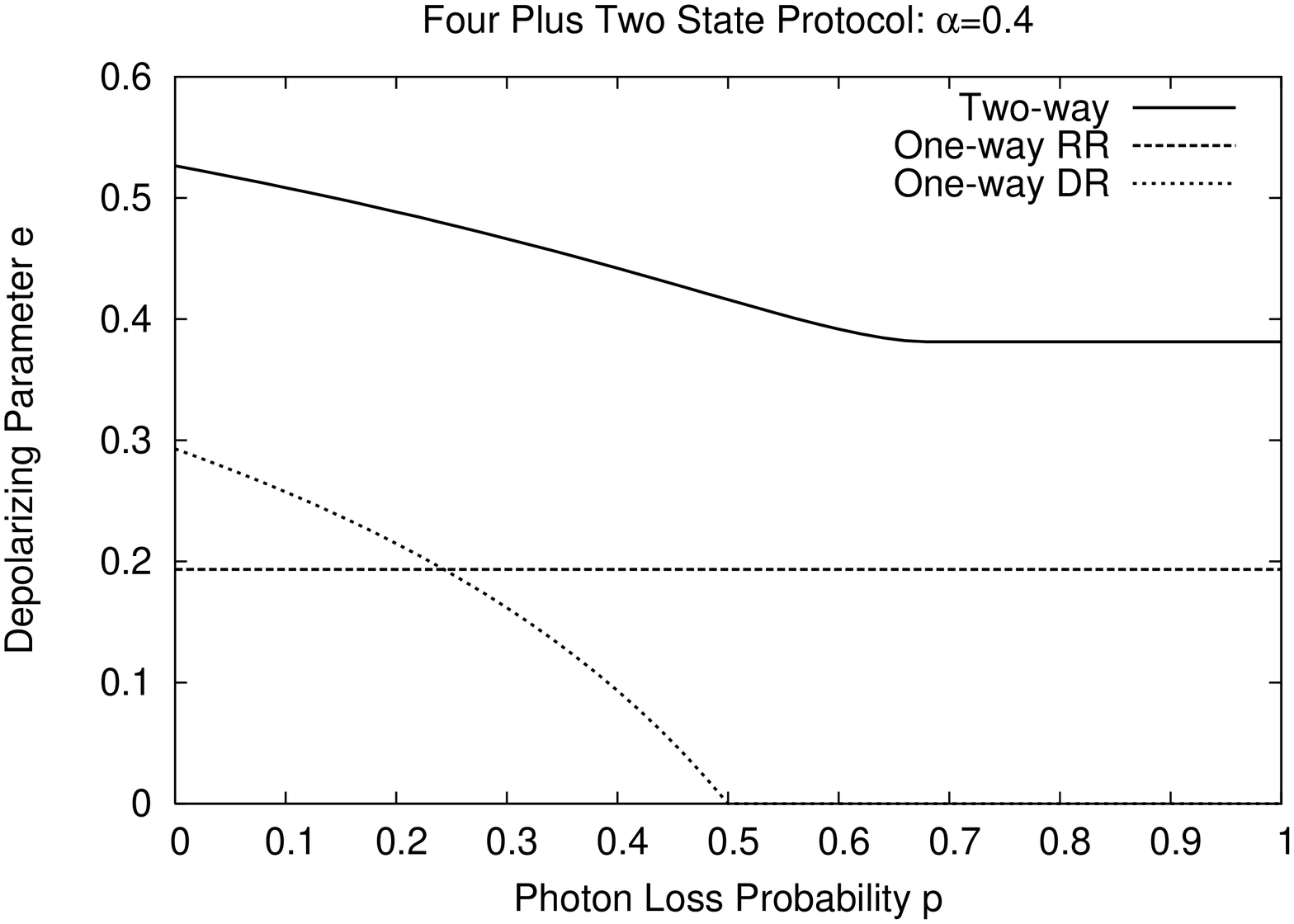}
\end{center}
\caption{Upper bound on the depolarising rate $e$ as a function of
the photon loss probability $p$ for the qubit-based
four-plus-two-state QKD protocol. Two-way classical
post-processing is illustrated with a solid line, while one-way
classical post-processing is represented with a dashed line
for RR, and with a dotted line for DR. The cases $\theta=0$
and $\theta=\pi/8$ provide the same results.
\label{fptwo-state-fig}}
\end{figure}

The resulting upper bound on the depolarising rate $e$ is
illustrated in Fig.~\ref{fptwo-state-fig} for the cases
$\alpha=0.2$ and $\alpha=0.4$.
The QBER is given by
\begin{equation}
QBER=\frac{e+(1-e)[1+(\alpha^2-\beta^2)^2]\sin^2\theta}{2\big[e+(1-e)
[2(\alpha^4+\beta^4)\sin^2\theta+4\alpha^2\beta^2\cos^2\theta]\big]}.
\end{equation}
In the case of two-way classical post-processing, the maximum
tolerable value of $e$ shown in Fig.~\ref{fptwo-state-fig} starts
decreasing as the losses in the channel increase, and, at some
point, it becomes constant independently of $p$. Interestingly,
the value of $p$ where this inflexion occurs, corresponds to the
point where Eve can discriminate unambiguously between the two
states in the set $\{|\varphi_k\rangle\}$, with $k=0,1$, or
between those states in the set $\{|\varphi_{\bar{k}}\rangle\}$,
with $k=0,1$. This happens when $p=1-2\alpha^2$.

\subsection{Three-state protocol}

This QKD scheme requires Alice sending to Bob one of the following
three quantum states $\ket{0},\ket{1},\ \text{and}\ \ket{+}$
\cite{three1,three2,three3,fred}. On the receiving side, Bob
projects each incoming signal onto one of the two possible bases
used in the four-state protocol (see Sec.~\ref{four_sec}),
together with a projection onto the vacuum state $\ket{vac}$.

\begin{figure}
\begin{center}
\includegraphics[angle=-90,scale=.32]{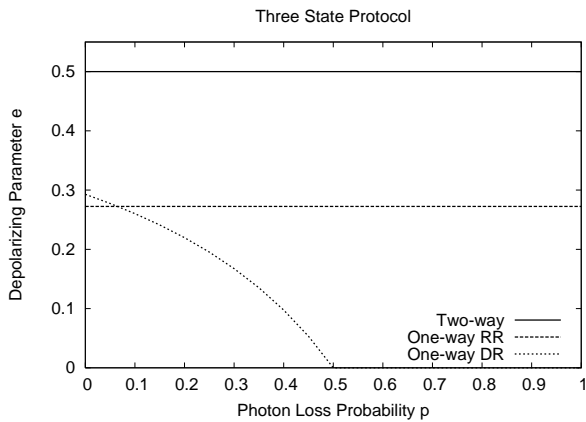}
\end{center}
\caption{Upper bound on the depolarising rate $e$ as a function of
the photon loss probability $p$ for the three-state QKD protocol.
Two-way classical post-processing is illustrated with a solid
line, while one-way classical post-processing is represented with
a dashed line for RR, and with a dotted line for DR. The cases
$\theta=0$ and $\theta=\pi/8$ provide the same
results.\label{fred-state-fig}}
\end{figure}

The resulting upper bound on the depolarising rate $e$ is
illustrated in Fig.~\ref{fred-state-fig}.
The QBER has now the following form
\begin{equation}
QBER=\frac{1}{2}\bigg[1+(1-e)(\sin^2\theta-\cos^2\theta)\bigg].
\end{equation}
For the quantum channel given by Eq.~(\ref{channel_eq}), and
assuming $\theta=0$ or $\theta=\pi/8$, the maximum value of $e$
tolerated by the three-state protocol coincides with the
four-state protocol for the cases of two-way and one-way
post-processing with DR.

\subsection{Trine protocol}

In the trine protocol \cite{joe04}, Alice selects, at random and
independently each time, a qubit in one of the following three
states: $\ket{0}, 1/2\ket{0}+\sqrt{3}/2\ket{1}$, and
$1/2\ket{0}-\sqrt{3}/2\ket{1}$, and sends it to Bob. Each
received signal is measured by Bob with a POVM defined by the
following operators: $B_0=2/3\ket{1}\bra{1}$,
$B_i=2/3\ket{\psi_i}\bra{\psi_i}$, with $i=1,2$, and where
$\ket{\psi_i}=\sqrt{3}/2 \ket{0}+ (-1)^i 1/2 \ket{1}$, and
$B_{vac}=\ket{vac}\bra{vac}$.

For the quantum channel given by Eq.~(\ref{channel_eq}), and
assuming $\theta=0$ or $\theta=\pi/8$, it turns out that the
maximum value of $e$ tolerated by this scheme, both for two-way
and one-way post-processing, coincides with the four-state
protocol (see Fig.~\ref{four-state-fig}). The QBER, however, is
given by
\begin{equation}
QBER=\frac{2e+4(1-e)\sin^2\theta}{3+e+2(1-e)\sin^2\theta}.
\end{equation}

\subsection{Ac\'{\i}n-Massar-Pironio protocol}

\begin{figure}
\begin{center}
\includegraphics[angle=-90,scale=.32]{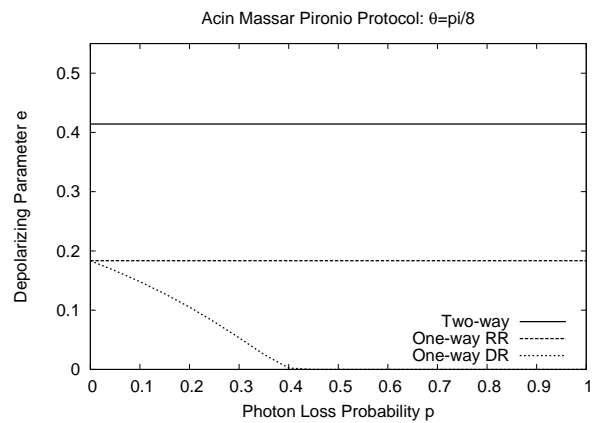}
\end{center}
\caption{Upper bound on the depolarising rate $e$ as a function of
the photon loss probability $p$ for the QKD protocol in
Ref.~\cite{ac06} when $\theta=\pi/8$. Two-way classical
post-processing is illustrated with a solid line, while one-way
classical post-processing is represented with a dashed line for
RR, and with a dotted line for DR. \label{acin1-state-fig}}
\end{figure}
In this scheme, Alice sends to Bob one of the following six
states: $\ket{+},\ket{-},1/\sqrt{2}(\ket{0}\pm{}i\ket{1})$, and
$1/\sqrt{2}\ket{0}\pm(1-i)/2\ket{1}$ \cite{ac06,noteac06}. On the
receiving side, Bob measures each incoming signal with one of two
possible measurements corresponding to the bases
$1/\sqrt{2}(\ket{0}\pm{}e^{-i\phi}\ket{1})$ with
$\phi=\{\pi/4,-\pi/4\}$, that he selects at random and
independently for each signal, together with a projection onto
the vacuum state $\ket{vac}$.

When $\theta=0$, the maximum value of $e$ tolerated by this
protocol, both for two-way and one-way post-processing, coincides
with the four-state protocol (see Fig.~\ref{four-state-fig}). The
case $\theta=\pi/8$ is illustrated in Fig.~\ref{acin1-state-fig}.

The QBER is now given by
\begin{equation}
QBER=\frac{1}{2}\bigg[(1-e)\sin^2\theta+e\bigg].
\end{equation}

\newpage

\bibliographystyle{apsrev}
\bibliographystyle{apsrev}

\begin{thebibliography}{99}
\bibitem{note_neumark} Using the Schmidt decomposition we have that the state
$\ket{\psi_{source}}_{AB}$ can be written as
$\ket{\psi_{source}}_{AB}=\sum_{k=0}^1
\sqrt{q_k}\ket{u_k}_A\ket{v_k}_B$ for some orthogonal states
$\ket{u_k}_A=\sum_{i=0}^{N-1}\ c^k_i \ket{\alpha_i}_A$ and where
$\ket{v_k}_B\in\mathcal{H}^B_2$. The orthogonal basis
$\ket{\alpha_i}_A$ can always be selected such as
$\ket{\alpha_i}_A=\ket{0}_{A^{\prime}}\ket{\beta_i}_{A^{\prime\prime}}$
for $i=0,...,\lfloor{}\frac{N-1}{2}\rfloor$, and
$\ket{\alpha_i}_A=\ket{1}_{A^{\prime}}
\ket{\beta_{(i-\lfloor{}\frac{N-1}{2}\rfloor-1)}}_{A^{\prime\prime}}$
for $i=\lfloor{}\frac{N-1}{2}\rfloor+1,...,N-1$, with
$A^{\prime}$ and $A^{\prime\prime}$ denoting two auxiliary
systems under Alice's control, and where
$\ket{\beta_i}_{A^{\prime\prime}}$ satisfy
$\bra{\beta_i}\beta_j\rangle_{A^{\prime\prime}}=\delta_{ij}$. On
the other hand, the states $\ket{u_k}_A$ can always be obtained
from the orthogonal states
$\ket{k}_{A^{\prime}}\ket{\beta_0}_{A^{\prime\prime}}$ by means
of a unitary transformation $U_{A^{\prime}A^{\prime\prime}}$,
{\it i.e.}, $\ket{u_k}_A =
U_{A^{\prime}A^{\prime\prime}}\ket{k}_{A^{\prime}}\ket{\beta_0}_{A^{\prime\prime}}$.
With this notation, we can write the initial state prepared by
Alice as $(U_{A^{\prime}A^{\prime\prime}}\otimes\openone_B)
\ket{\psi_{source}}_{A^{\prime}B}\ket{\beta_0}_{A^{\prime\prime}}$
with $\ket{\psi_{source}}_{A^{\prime}B}=\sum_{k=0}^{1}
\sqrt{q_k}\ket{k}_{A^{\prime}}\ket{v_k}_B\in
\mathcal{H}^{A^{\prime}}_2\otimes\mathcal{H}^B_2$. The
measurement process in the basis $\ket{\alpha_i}_A$ can then be
described as: Alice first prepares the signal state
$\ket{\psi_{source}}_{A^{\prime}B}$, and then she measures her
subsystem $A^{\prime}$ with the POVM operators
$A_i={}_{\phantom{1}A^{\prime\prime}}\bra{\beta_0}U_{A^{\prime}A^{\prime\prime}}^\dag\ket{\alpha_i}_A
\bra{\alpha_i}U_{A^{\prime}A^{\prime\prime}}\ket{\beta_0}_{A^{\prime\prime}}$
with $i=0,...,N-1$.
\bibitem{partialtrans} Given an operator
$\rho_{AB}\in \mathcal{H}^A\otimes\mathcal{H}^B$ and an
orthonormal basis $\{|\beta_i\rangle_B\}\in \mathcal{H}^B$, with
$i=1,...,N$, the partial transposed of $\rho_{AB}$ with respect
to subsystem $B$ in that basis is defined as
\begin{equation}
\rho_{AB}^{\Gamma}=\sum_{i,j=1}^N
{}_{\phantom{1}B}\langle{}\beta_i|\rho_{AB}|\beta_j\rangle_B\
                       |\beta_j\rangle_B\langle{}\beta_i|
                       \end{equation}
In the same way, one can also define the partial transposed of
$\rho_{AB}$ with respect to subsystem $A$.
\bibitem{tobi_note} The operator $\rho_{\text{fix}}$ can always be
expressed as $\rho_{\text{fix}}=\sum_{kl \in I} c_{kl} S_{kl}$
because of the following reason: Suppose, for instance, that a
given set of measurement operators $\{ A_i \otimes B_j \}$ do not
form a set of operator basis elements. Then, one can find the
minimal set of linear independent measurement operators and then
apply the Gram-Schmidt orthogonalization to these elements in
order to obtain a minimal set of linear independent operator basis
elements. Moreover, from the original data $p_{ij}$, it is
straightforward to compute the observed probability distribution
for these new operators. As a result, one finds a fixed part
$\rho_{\text{fix}}$ of the form $\rho_{\text{fix}}=\sum_{kl \in
I} c_{kl} S_{kl}$.
\bibitem{metodo_alternativo} Alternatively to this method, the
equality constraints given by Eq.~(\ref{cond_coeff}) can also be
included directly in the LMI constraint of the SDP. Each of these
constraint can be represented by means of two inequality
constraints as follows: $\sum_{kl} a_{ik} b_{jl}
x_{kl}-p_{ij}\geq 0$, and $-(\sum_{kl} a_{ik} b_{jl}
x_{kl}-p_{ij})\geq 0$. This approach, however, increases the
number of objective variables to be considered.
\bibitem{notefpt} In its original proposal, the
four-plus-two-state protocol was meant to operate with weak
coherent pulses as signal states \cite{hut95}. Here we refer to
the single-photon version of this protocol.
\bibitem{noteac06} The main motivation behind this protocol
is that it can be proven to be secure also against an hypothetical
Eve only limited by the no-signalling principe \cite{ac06}. In
its original proposal, Alice and Bob share a noisy quantum channel
that distributes pairs of qubits in the maximally entangled state
$\ket{\psi}_{AB}=1/\sqrt{2}(\ket{00}+\ket{11})$. These pairs are
then measured by Alice and Bob. Here, however, we consider a
prepare and measure version of this scheme with Alice being the
source of the signal states that are sent to Bob.
\end{thebibliography}

\end{document}